\pdfoutput=1
\documentclass[reprint,12pt,onecolumn,notitlepage,nofootinbib,floatfix]{revtex4-2}
\usepackage{amssymb}
\usepackage{amsmath}
\usepackage{amsfonts} 
\usepackage{bm}
\usepackage{color} 
\usepackage{subfigure}
\usepackage{graphicx}
\usepackage[bookmarks=false]{hyperref} 
\hypersetup{pdfstartview=FitH,pdfhighlight=/O,colorlinks=true}

\allowdisplaybreaks

\usepackage{epsfig} 
\usepackage{epstopdf}
\usepackage{latexsym}
\usepackage{psfrag}   
\usepackage{subfigure}  
\usepackage{booktabs}  
\usepackage{braket}
\usepackage{mathtools}
\usepackage{textcomp}
\usepackage{ifthen}
\usepackage{tensor}
\usepackage{protosem} 
\usepackage{wasysym}

\usepackage[toc]{appendix}
\usepackage{color,soul} 
\usepackage{datetime}

\definecolor{darkblue}{rgb}{0.2, 0, 0.8}
\definecolor{darkgreen}{rgb}{0.2, 0.71, 0}
\definecolor{awesome}{rgb}{1.0, 0.13, 0.32}
\definecolor{cadmiumred}{rgb}{0.89, 0.0, 0.13}
\definecolor{dukeblue}{rgb}{0.0, 0.0, 0.61}

\newcommand{\bea}{\begin{eqnarray}}
\newcommand{\eea}{\end{eqnarray}}
\newcommand{\ba}{\begin{eqnarray}}
\newcommand{\ea}{\end{eqnarray}}

\newcommand{\beq}{\begin{equation}}
\newcommand{\eeq}{\end{equation} }
\newcommand{\beqa}{\begin{eqnarray}}
\newcommand{\eeqa}{\end{eqnarray}}
\newcommand{\beqar}{\begin{eqnarray*}}
\newcommand{\eeqar}{\end{eqnarray*}}

\renewcommand{\href}[2]{#2}

\begin{document}

\title{Large fluctuations in the Sky}

\author{Sayantan Choudhury}
\email{sayanphysicsisi@gmail.com, sayantan\_ccsp@sgtuniversity.org, sayantan.choudhury@nanograv.org (Corresponding Author)}
\affiliation{Centre For Cosmology and Science Popularization (CCSP),
        SGT University, Gurugram, Delhi- NCR, Haryana- 122505, India }

\date{\today}

\begin{abstract}  \vskip 0.2in 

Renormalization of quantum loop effects generated from large fluctuations is a
hugely debatable topic of research these days which rules out 
the Primordial Black Hole (PBH) formation within the framework of single-field inflation. In this article, we briefly discuss that the correct implementation of regularization, renormalization, and resummation techniques in a setup described by an ultra-slow-roll phase sandwiched between two slow-roll phases in the presence of smooth or sharp transitions can lead to a stringent constraint on the PBH mass (i.e. ${\cal O}(10^{2}{\rm gm}$)), which we advertise as a new {\it No-go theorem}. Finally, we will give some of the possible way-outs using which one can evade this proposed {\it No-go theorem} and produce solar/sub-solar mass PBHs.

\vskip 0.2in
\centering
\noindent {\it \footnotesize Essay written and received honorable mention for the Gravity Research Foundation 2024 Awards for
Essays on Gravitation}

\end{abstract}   

\maketitle 

\newpage

\section{Introduction}

The investigation of Primordial Black Hole (PBH) generation \cite{Zeldovich:1967lct,Hawking:1974rv,Carr:1974nx,Carr:1975qj,Chapline:1975ojl,Carr:1993aq,Kawasaki:1997ju,Yokoyama:1998pt,Kawasaki:1998vx,Rubin:2001yw,Khlopov:2002yi,Khlopov:2004sc,Saito:2008em,Khlopov:2008qy,Carr:2009jm,Choudhury:2011jt,Lyth:2011kj,Drees:2011yz,Drees:2011hb,Choudhury:2013woa,Ezquiaga:2017fvi,Kannike:2017bxn,Hertzberg:2017dkh,Pi:2017gih,Gao:2018pvq,Dalianis:2018frf,Cicoli:2018asa,Ozsoy:2018flq,Byrnes:2018txb} from the inflationary paradigm has gained attention due to a number of recent discoveries. These include the scalar field models, both canonical and non-canonical, some of which include abrupt or smooth transitions from a slow-roll (SR) phase to an ultra-slow-roll (USR) phase. The basic explanation for the production of PBHs in the early phases of the universe is an augmentation of fluctuation at a specific scale brought about by a mechanism related to the inflaton field's motion on the flat potential. Here, the inflaton field travels along the otherwise flat potential and comes into contact with one or more tiny, transient bumps \cite{Mishra:2019pzq,ZhengRuiFeng:2021zoz}. This causes the field to fluctuate sufficiently to leave an impression on the corresponding scale. Even while adding a bump to an otherwise flat potential could make sense phenomenologically, theoretical rationale forces this to be considered a liability. On the other side, an increase in volatility that could lead to PBH production later on could result from a transition from SR to USR. Without having to know the inflaton potential, quantum phenomena can be studied in this framework in a model-independent manner. Attempts have been made to integrate the different physics of these models within the framework of Effective Field Theory (EFT) for single field inflation, with an effective action that is valid below a given UV cut-off scale \cite{Weinberg:2008hq,Cheung:2007st,Choudhury:2017glj,Delacretaz:2016nhw}. We decide to carry out our analysis inside EFT's borders.

Recently the authors of refs. \cite{Kristiano:2022maq,Kristiano:2023scm} argue that PBH creation via single-field inflation is excluded. The basis for their conclusions is the claim that, on a large scale, a one-loop adjustment to the tree-level power spectrum is incredibly huge. This compelling argument primarily rests on the observation that, in addition to logarithmic divergent effects, there are quadratic divergent contributions at the one-loop primordial power spectrum result. A persistent difficulty has been producing PBHs with masses large enough, ${\cal O}(M_{\odot})$ (solar mass), while taking quantum loop corrections into account in the scalar power spectrum. 
The lack of subtleties in the renormalization and resummation processes in these methods, which are essential for an inflationary framework, has been a major drawback, though. A recent model-independent method was used to contradict this claim in refs.\cite{Riotto:2023hoz,Riotto:2023gpm}, demonstrating that short-scale loop effects have no effect on the large-scale primordial power spectrum. Such a contradiction puts the subject in the limelight and to date, a lot of efforts have been made either to support or refute the corresponding argument \cite{Kristiano:2022maq,Kristiano:2023scm,Riotto:2023hoz,Riotto:2023gpm,Firouzjahi:2023aum,Firouzjahi:2023ahg,Firouzjahi:2023bkt,Choudhury:2023vuj,Choudhury:2023jlt,Choudhury:2023rks,Motohashi:2023syh,Franciolini:2023lgy,Cheng:2023ikq,Tasinato:2023ukp,Tasinato:2023ioq, Iacconi:2023ggt,Davies:2023hhn}. By correct implementation of regularization, renormalization, and resummation techniques in this article, we are going to explicitly justify that the PBH formation from single field inflationary paradigm is ruled out considering the quantum loop effects at the one-loop corrected primordial power spectrum obtained from the EFT framework \cite{Choudhury:2023vuj,Choudhury:2023jlt,Choudhury:2023rks}. In support of this argument, we further propose a new {\it No-go theorem} on the PBH mass, which only allows the generation of very small PBHs having mass, ${\cal O}(10^{2}{\rm gm})$ and rules out the possibility for generating large mass (solar or sub-solar mass) PBHs within the present framework. Last but not least, in the end, we are going to provide the best possible way-outs to evade this proposed {\it No-go theorem} on the PBH mass \cite{Choudhury:2023hvf,Choudhury:2023kdb,Choudhury:2023hfm,Choudhury:2023fwk,Choudhury:2024one,Bhattacharya:2023ysp,Choudhury:2023fjs,Choudhury:2024dei}, as an immediate outcome of which one can safely in a correct framework allow generation of large mass PBHs within the single field inflationary paradigm.

\section{The ultimate three-phase scenario for PBH formation}\label{sec2}

The fundamental notion in the framework under discussion is to start with a model-independent, effective action that is valid below the UV cut-off scale. In addition, symmetry imposes constraints on the structure of the EFT action. We may provide a constraint on the speed of sound ($c_s$) using this approach, which is defined in terms of EFT parameters. When utilizing the St$\ddot{u}$ckelberg approach, which basically consists of scalar perturbation known as Goldstone modes, we choose the unitary gauge. One could consider integrating the Goldstone mode into the non-linear sigma model framework, which is also viewed as the UV-completed form of the linearized gauge symmetry, in a similar manner to the Standard Model Higgs sector.

Let us start with the following EFT action:
\bea
 S&=&\displaystyle\int d^{4}x \sqrt{-g}\left[\frac{M^2_{pl}}{2}R+M^2_{pl} \dot{H} g^{00}-M^2_{pl} \left(3H^2+\dot{H}\right)+{\cal F}\left(\delta g^{00}, \delta K^{\mu\nu},\cdots\right)\right].
	\eea
Here the last term physically signifies all possible contributions from the small perturbations in quasi de Sitter background. Here, $K_{\mu\nu}$ defines the extrinsic curvature at a constant time slice and $H$ represents the Hubble parameter. In this setup the Goldstone mode ($\pi(t, {\bf x})$) transforms under the time diffeomorphism symmetry $t\rightarrow t+ \xi^{0}(t,{\bf x})$ as, 
$\pi(t, {\bf x})\rightarrow\pi(t, {\bf x})-\xi^{0}(t,{\bf x}),$
   where the local parameter is $\xi^{0}(t,{\bf x})$ and gauge is fixed by imposing $\pi(t,{\bf x})=0$. These Goldstone mimics the role of scalar perturbation in this description. Further ignoring the contribution from the gravity and Goldstone mode mixing above the characteristic scale, $E>E_{\rm mix}=\sqrt{\dot{H}}$. Imprints of the scalar perturbations at the comoving scale can be further described in terms of the curvature perturbation variable ($\zeta$) through the linear mapping in terms of the Goldstone modes ($\pi$) in the corresponding second-order perturbed EFT action. Once this connection is established, the curvature perturbation variable ($\zeta$) is described in terms of a second-order differential equation, which is commonly known as {\it Mukhanov Sasaki equation} physically describing a system of oscillators having a time-dependent frequency in momentum space. 
   
   To explain the generation of PBHs, in this article, we primarily focus on the three-phase scenario, which we are identifying to be an {\it ultimate} framework in this context as it captures all the necessary ingredients to generate large amplitude fluctuations. In a more precise language, this can be easily done by considering an ultra-slow roll (USR) phase sandwiched between two slow roll (SRI and SRII) phases. At the end of the SRII phase inflation ends when the number of e-foldings $N$ reaches the magic number $60$, which is necessarily required to accomplish inflation successfully in the present context of the discussion. Furthermore, depending on the behaviour of the transition from SRI to USR and USR to SRII phases on the technical ground of the present model-independent EFT setup one can further parameterize the second slow-roll parameter $\eta=\epsilon-\frac{1}{2}\frac{d\ln{\epsilon}}{dN}$ (where the first slow-roll parameter $\epsilon=-\frac{d\ln H}{dN}$) by the following generic fashion:
   \bea
\eta(N) = \eta_{\rm I}(N\leq N_{s}) + f(N-N_{s})\eta_{\rm II}(N_{s}\leq N\leq N_{e}) + f(N-N_{e})\eta_{\rm III}(N_{e}\leq N\leq N_{\rm end}),\quad
\eea
\begin{figure*}[ht!]
    	\centering
    \subfigure[]{
      	\includegraphics[width=7.9cm,height=6cm]{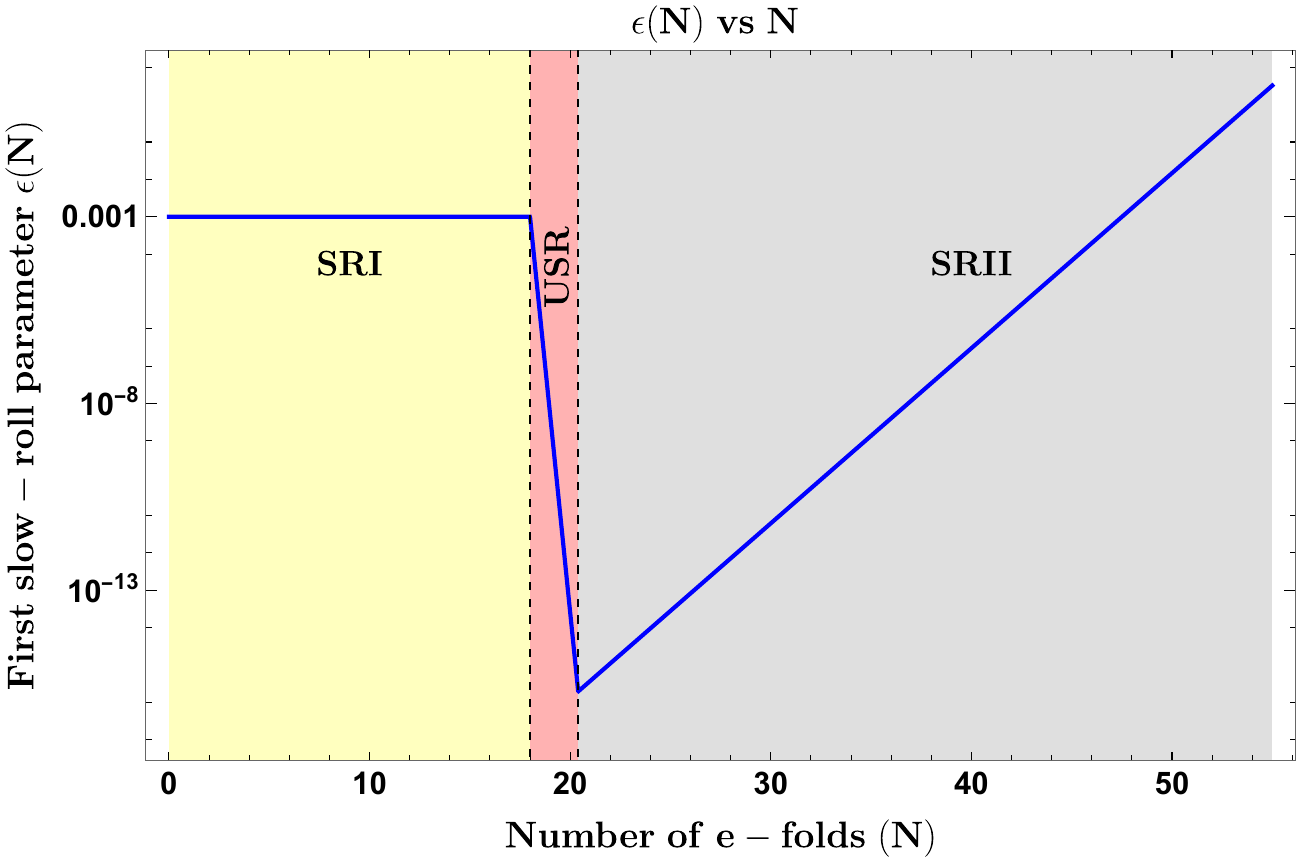}
        \label{epsilon}
    }
    \subfigure[]{
        \includegraphics[width=7.9cm,height=6cm]{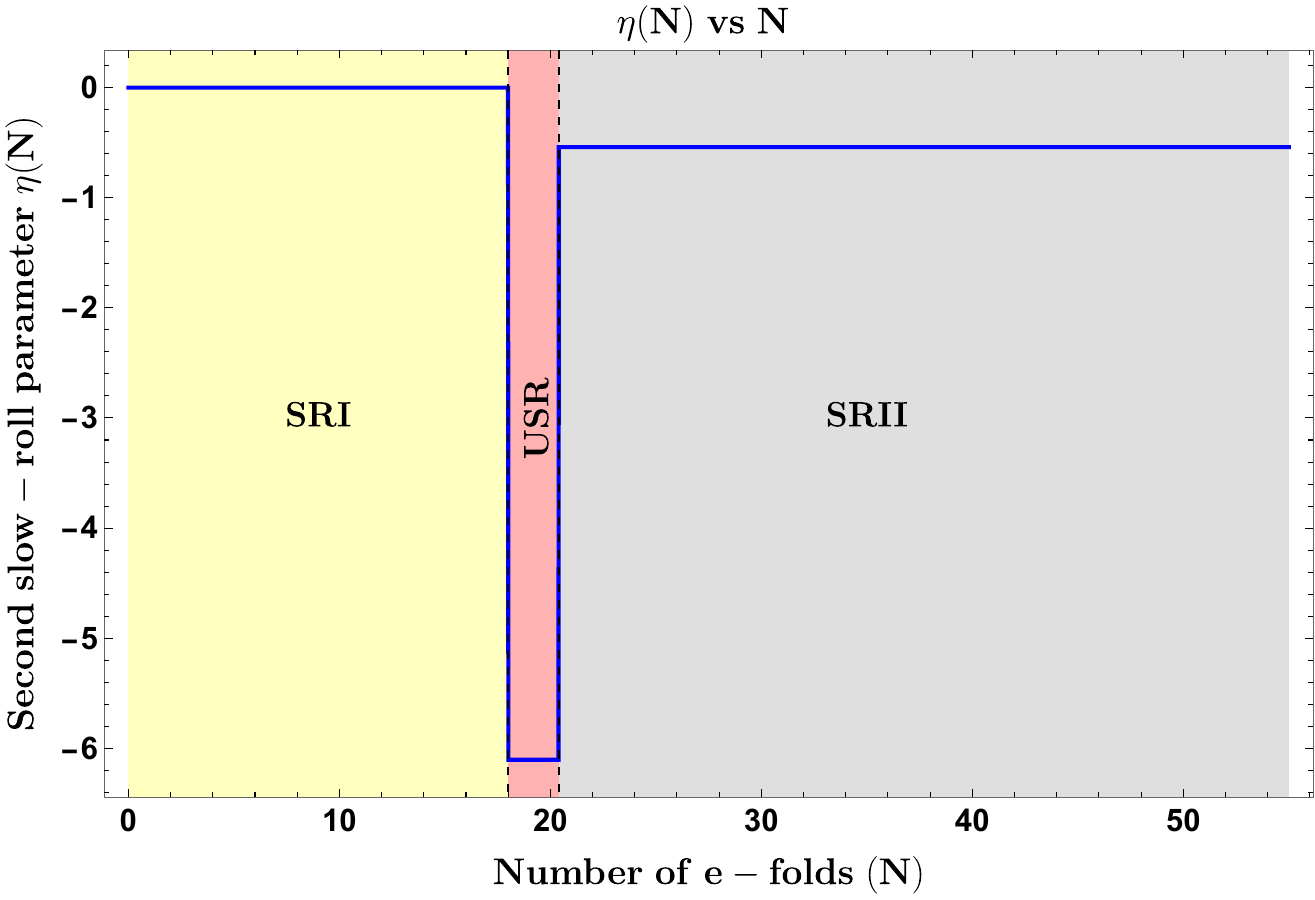}
        \label{eta}
    }
    	\caption[Optional caption for list of figures]{Behaviour of the \ref{epsilon} first slow-roll parameter $\epsilon({ N})$ and \ref{eta} second slow-roll parameter $\eta({ N})$ in presence of a USR phase as a function of the e-foldings $N$ for sharp transition.} 
    	\label{dyn}
    \end{figure*}
where the functions characterizing the transitions are given by, $f(N-N_a)={\rm tanh}\left(\frac{N-N_a}{\Delta N}\right)$ with  $a=s\;$ (SRI$\rightarrow$ USR),$e\;$ (USR$\rightarrow$ SRII). Here $\Delta N$ represents the width of the USR phase and technically this specific feature of the function actually describes the smooth transitions in the present context. In the limiting case $\Delta N\rightarrow 0$ corresponds to $f(N-N_a)=\Theta(N-N_a)$, which are sharp transitions described in terms of Heaviside Theta functions. Additionally, it is important to note that, $\eta_{\rm I},\;\eta_{\rm II},\;\eta_{\rm III}$ describes constants appearing in SRI, USR, and SRII phases respectively. Utilizing this parametrization one can further construct, $\epsilon(N) = \eta\left[1-\left(1-\frac{\eta}{\epsilon_{i}}\right)e^{2\eta (N-N_i)}\right]^{-1}$, where $(\epsilon_{i},N_{i})$ represents a preferred initial condition. In figure \ref{epsilon} and \ref{eta} we have depicted the features of the first and second slow-roll parameters with respect to the number of e-foldings $N$ in the presence of sharp transition ($\Delta N\rightarrow 0$). At the boundaries of the SRI to USR and USR to SRII transitions the behaviour will be comparatively smoother when we have non-negligible width ($\Delta N$) of the USR phase. 

Further utilizing the above-mentioned facts demonstrated in the consecutive three-phase scenario the generic solution for the curvature perturbation modes in momentum space from the {\it Mukhanov Sasaki equation} can be further written as:
\bea
 &&\zeta_{\bf k}(\tau)=\left(\frac{ic_sH}{2M_{ pl}\sqrt{\epsilon_{P}(\tau)}}\right)\frac{1}{(c_sk)^{3/2}}\left[\alpha^{(P)}_{{\bf k}}\left(1+ikc_s\tau\right)\; e^{-ikc_s\tau}-\beta^{(P)}_{{\bf k}}\left(1-ikc_s\tau\right)\; e^{ikc_s\tau}\right]\forall P,\quad\quad\quad
 \eea
where $P$ represents the phase identification index which is given by $P= 1 ({\rm SRI})$, $2 ({\rm USR})$, $3 ({\rm SRII})$. Additionally, it is important to note that, in the USR and SRII phases we have, $\epsilon_{\rm USR}=\epsilon_{\rm SRI}(\tau_s/\tau)^6$ and $\epsilon_{\rm SRII}=\epsilon_{\rm SRI}(\tau_s/\tau_e)^6$. Here $\alpha^{(P)}_{{\bf k}}$ and $\beta^{(P)}_{{\bf k}}$ are the Bogoliubov coefficients for these mentioned three phases. For a given initial choice of vacuum, which is commonly described in terms of the {\it Bunch Davies} states in the SRI phase one can further fix the expressions for both of the Bogoliubov coefficients in the USR and SRII phases (the newly shifted vacua in these two phases are away from {\it Bunch Davies} initial states) by utilizing the continuity and differentiability conditions of the curvature perturbation modes. Last but not least, in this three-phase construction we also assume that at SRI to USR and USR to SRII transition points the effective sound speed $c_s$ is approximately the same and parametrized as, $c_s=1\pm \delta$, where $\delta$ is very small tuning factor. On the other hand within SRI, USR, and SRII phases the effective sound speed is described by its pivot scale value $c_{s,*}$ which is smaller than the value at transition points, i.e. $c_{s,*}<c_s$.

\section{Regularized-Renormalized-Resummed power spectrum}

\begin{figure*}[ht!]
    	\centering
    \subfigure[]{
      	\includegraphics[width=7.9cm,height=6cm]{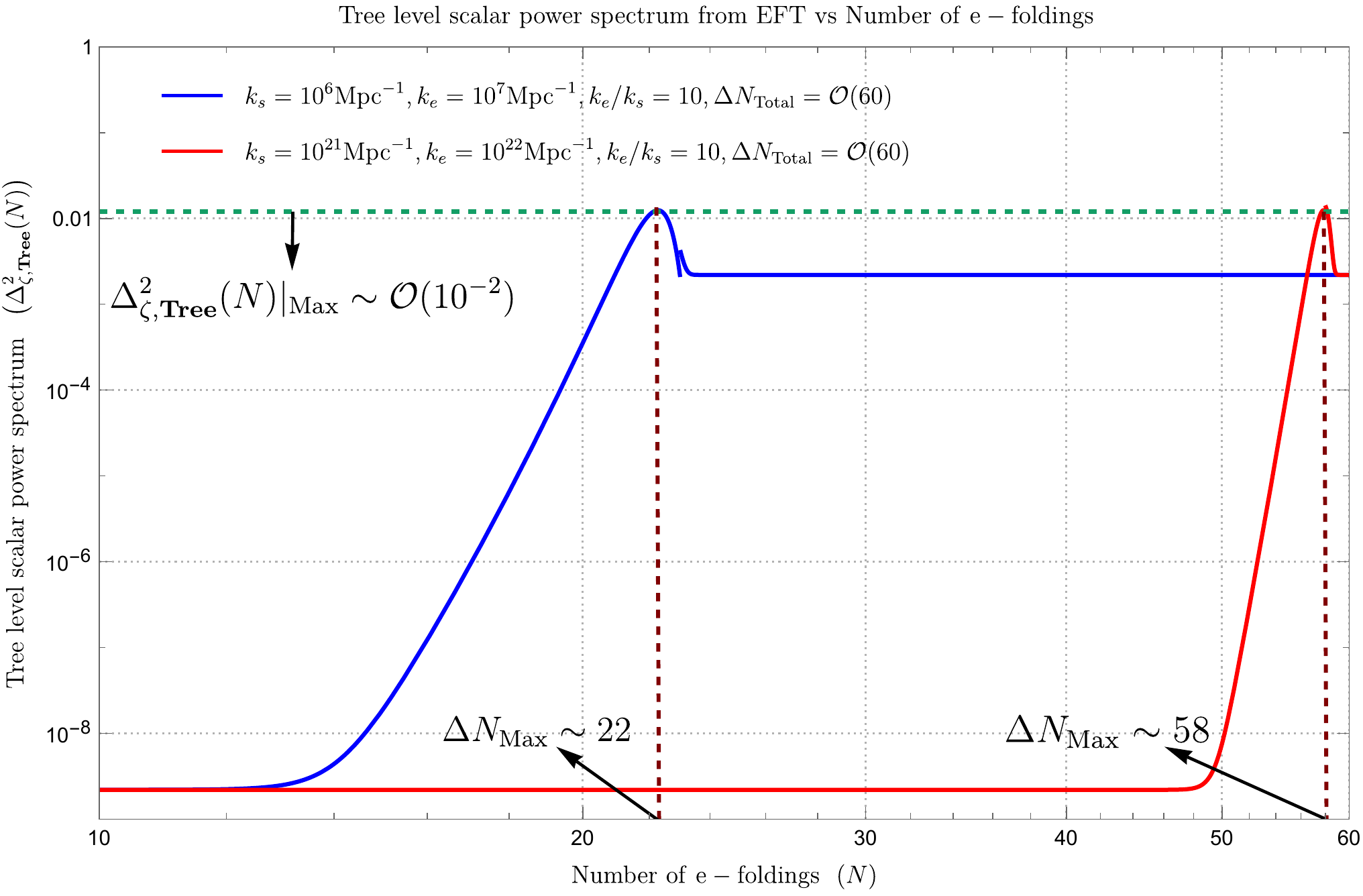}
        \label{Tree}
    }
    \subfigure[]{
        \includegraphics[width=7.9cm,height=6cm]{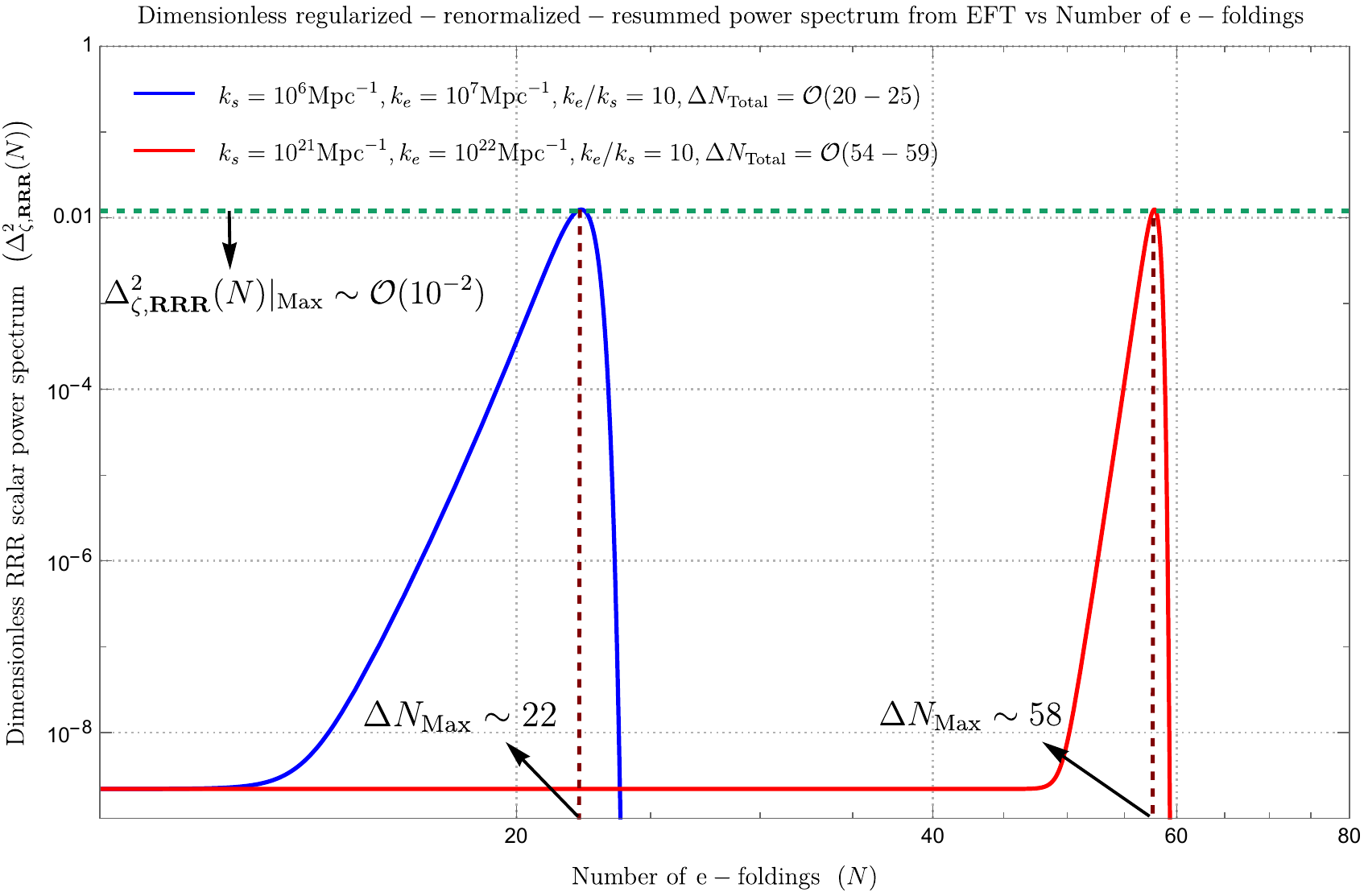}
        \label{RRR}
    }
    	\caption[Optional caption for list of figures]{Behaviour of the \ref{Tree} tree-level and \ref{RRR} regularized-renormalized-resummed power spectrum with respect to the e-foldings $N$ for sharp/smooth transition.} 
    	\label{dyn}
    \end{figure*}

Before discussing the physical impacts of the quantum loop effects on the primordial power spectrum for the comoving scalar curvature perturbation, let us first quantify the tree-level contribution, which can be computed in terms of the physical modes computed in the three-phase scenario by the following expression: 
\bea \label{tree}\Delta^{2}_{\zeta,{\bf Tree}}(k)=\Delta^{2}_{\zeta,{\bf SRI}}(k)\times\Bigg\{1+\left(\frac{k_e}{k_s }\right)^{6}\bigg[f(k-k_s)\left|\alpha^{(2)}_{\bf k}-\beta^{(2)}_{\bf k}\right|^{2}+f(k-k_e)\left|\alpha^{(3)}_{\bf k}-\beta^{(3)}_{\bf k}\right|^{2}\bigg]\Bigg\},\quad\quad\eea
which describes both smooth and sharp transition scenarios depending on the behaviour of the function $f(k-k_a)\forall a$ as described before. In the above expression, the first term corresponds to the power spectrum in the SRI phase, which is given by, $\Delta^{2}_{\zeta,{\bf SRI}}(k)=\left(\frac{H^{2}}{8\pi^{2}M^{2}_{ pl}\epsilon c_s}\right)_*$. It is important to note that here the expression in the SRI phase is quantified at the pivot scale. The tree-level spectrum with respect to the e-foldings is depicted in figure \ref{Tree}. Here, it is important to note that the tree-level power spectrum contains divergences that must be treated with renormalization methods designed for curved spaces \cite{Agullo:2009vq,Agullo:2011qg,Ferreiro:2022ibf,Pla:2024xsv}.  

Further to describe the quantum loop effects, let us consider the interaction term $\zeta^{'}\zeta^{2}$ as appearing in the third-order perturbed action for the comoving curvature perturbation, which has a direct impact at the lowest order, i.e., at the one-loop level computation. By explicitly making use of the well-known Schwinger-Keldysh formalism within the framework of primordial cosmology, one can compute the impact of this contribution in the previously introduced three-phase scenario. As an outcome, the contribution of the previously mentioned cubic interaction term becomes large at the USR phase due to its large amplitude enhancement. From the other two phases (SRI and SRII) at the one-loop level, we get negligible contributions. Other terms appearing in the third-order action become suppressed for all of the three phases in this computation. 

However, the large contribution coming from $\zeta^{'}\zeta^{2}$ in the USR phase from the one-loop computation gives rise to problematic quadratic UV and less harmful logarithmic IR divergence. This directly makes the use of perturbation theory questionable, using which it was recently claimed by various authors \cite{Kristiano:2022maq,Kristiano:2023scm} that the generation of PBH formation is ruled out within the framework of a single-field inflationary paradigm in the presence of sharp transitions. Here, it is important to note that the Heaviside Theta function is used in this context to describe the sharp transition. Later, a group of other authors \cite{Riotto:2023hoz,Riotto:2023gpm,Firouzjahi:2023aum,Firouzjahi:2023ahg,Firouzjahi:2023bkt} refuted this claim by explicitly showing that in the context of smooth transitions from SRI to USR and USR to SRII phases, the one-loop correction turns out to be suppressed, hence PBH formation from single field inflation is not ruled out. Many works have been done in this direction, either to support or refute the argument \cite{Kristiano:2022maq,Kristiano:2023scm,Riotto:2023hoz,Riotto:2023gpm,Firouzjahi:2023aum,Firouzjahi:2023ahg,Firouzjahi:2023bkt,Choudhury:2023vuj,Choudhury:2023jlt,Choudhury:2023rks,Motohashi:2023syh,Franciolini:2023lgy,Cheng:2023ikq,Tasinato:2023ukp,Tasinato:2023ioq,Iacconi:2023ggt,Davies:2023hhn}, which has given rise to a huge debate that is ongoing to date. Unfortunately, out of all of these works, not a single one has implemented the regularization-renormalization-resummation techniques correctly to give concrete technical proof to refute or support this argument. In refs. \cite{Firouzjahi:2023aum,Firouzjahi:2023ahg,Firouzjahi:2023bkt} the authors have shown that smooth transitions can be implemented by changing the strength of the Heaviside Theta function by introducing a parameter $h$, which further implies that $\Theta(k-k_a)$ is replaced by $h\Theta(k-k_a)$, where  $a=s\;$ (SRI$\rightarrow$ USR), $e\;$ (USR$\rightarrow$ SRII). They demonstrate the absence of quantum loop corrections at the lowest order (one-loop) by fixing a specific value of the parameter $h$ in this computation. Insertion of this parameter $h$ is similar to a kind of regularization that the authors in refs \cite{Firouzjahi:2023aum,Firouzjahi:2023ahg,Firouzjahi:2023bkt} have not pointed out explicitly. However, such an arrangement should work in all orders of perturbation theory. If in each order one requires a different value of regulator $h$ the purpose of such a regularization scheme is immediately defeated. This situation might be similar to the quantization of the four Fermi theory whose renormalization requires an infinite number of counterterms, which is not at all surprising as the latter is an effective theory that should not be quantized in principle. Hence, it is quite clearly explained in the above discussions that correct implementation and interpretation of regularization, renormalization, and resummation schemes are necessarily required to arrive at a physically justifiable conclusion in the present context. We have done various works to justify this issue in detail. See refs. \cite{Choudhury:2023vuj,Choudhury:2023jlt,Choudhury:2023rks} for more details. 

In this article, we are now summarizing the logical steps to follow to perform this computation and arrive at the final concrete conclusion on the formation of PBHs within the framework of a single-field inflationary framework. For more technical steps, please follow refs. \cite{Choudhury:2023vuj,Choudhury:2023jlt,Choudhury:2023rks}. These steps are appended below point-wise:
\begin{enumerate}
    \item \underline{\bf Regularization:} In our computation of the one-loop contributions from the primordial power spectrum for the scalar modes, we use a cut-off regularization scheme, which one can use in a trustworthy fashion in this context. To implement this method in all the one-loop integrals when we are integrating over internal momenta, instead of taking $0<k<\infty$, we divide the window of the integral into two parts: $k_{\rm IR}<k<k_{\rm INT}$ and $k_{\rm INT}<k<k_{\rm UV}$, where $k_{\rm UV}=\Lambda a(\tau)/c_s$, $k_{\rm IR}$ and $k_{\rm INT}$ represent the UV, IR cut-offs, and intermediate momentum scale. Here, $a(\tau)$ corresponds to the scale factor in the quasi de Sitter background. As an immediate consequence, the final result of one-loop contributions in the SRI, USR, and SRII phases contains the quadratic UV divergent term $\left(\Lambda/H\right)^2$ and the logarithmic divergent term $\ln\left(\Lambda/H\right)$. Here $\Lambda/H$ is a dimensionless quantity that contains the Hubble parameter. In this calculation, the coefficients appearing in front of these contributions are different in all three phases; the maximum contribution appears from the USR regime. Also, the expressions for these coefficients turn out to be approximately similar for both smooth and sharp transitions. In this context, out of the two types of divergences, UV divergences are more harmful and directly challenge the applicability of perturbation theory in the present context. 
    \item \underline{\bf Renormalization:} After performing the regularization in the correct fashion and obtaining the UV and IR divergent contributions, we need to get rid of them by applying the renormalization technique in this context. Technically, a correct renormalization scheme allows us to incorporate the counter-terms at the level of third-order action for the curvature perturbation. Here, the renormalized expression of the perturbation can be written in terms of its bare contribution as, $\zeta_{\bf R}=\sqrt{{\cal Z}}\zeta_{\bf B}$, where ${\cal Z}$ is the counter-term, which we need to explicitly determine to remove the divergences. After applying the correct {\it renormalization condition}, which is described in terms of fixing the renormalization scale at the Hubble scale and considering the fact that the renormalized primordial scalar power spectrum is at the CMB pivot scale is completely determined by the tree-level contribution of the SRI phase, we can very easily compute the expression for ${\cal Z}$. As an immediate consequence, the renormalized one-loop corrected primordial scalar power spectrum turns out to be completely free from the quadratic UV divergence. Also, the less harmful logarithmic IR divergent contribution appears with higher powers after applying the {\it renormalization condition}. In this discussion, all such higher powers of logarithmic contributions effectively mimic the higher-order loop diagrams having the same mathematical structure. Most importantly, we have checked that our findings are completely renormalization scheme-independent. This is a very important finding of our analysis, which allows us to correctly compute and physically interpret the applicability of our derived result in the perturbative regime. 

    \item \underline{\bf Resummation:} Finally, with the help of the computed renormalized primordial power spectrum for scalar modes we resum over all the possible logarithmic higher-order contributions with the help of the {\it Dynamical Renormalization Group} (DRG) technique, which commonly referred as {\it Exponentiation-influenced resummation}. This further implies that the DRG method allows for the sum over all repetitive loop diagrams, and the sum of all possible terms in the infinite series leads to a finite result due to its strict convergence at the horizon-crossing scales and super-horizon region. This outcome holds true for perturbative computations with any loop order that may accurately capture quantum effects. This outcome is applicable to a wider range of running momentum scales that are observationally feasible in the minuscule coupling domain, where the perturbative approximation is entirely valid within the context of the underlying EFT setup. 
    The resultant coarse-grained form of the power spectrum, with more softened logarithmic contributions than the renormalized version of the one-loop spectrum, is the most important output of the DRG resummed version of the one-loop corrected power spectrum, which is finally quantified as:
    \bea \Delta^2_{\zeta,{\bf RRR}}=\Delta^{2}_{\zeta,{\bf SRI}}(k)\times\exp\left(\sum_{\bf All~graphs} {\cal F}_{G}\right),\eea
where inside the exponential we have taken sum over all Feynman graphs appearing at loop-level. The final outcome of the resummed spectrum with respect to the e-foldings is depicted in figure \ref{RRR}.

\end{enumerate}
\section{ No-Go to Go}

From our analysis, we have found that the ratio of the momenta $k_e/k_s\sim {\cal O}(10)$ is necessarily required to maintain and validate the perturbative approximations in the resummed power spectrum. This ratio immediately restricts the span of the USR phase in terms of the number of e-foldings as, $\Delta N_{\rm USR}=N_e-N_s=\ln(k_e/k_s)\sim 2$. Additionally, we have found that to successfully accomplish inflation (i.e. to have $\Delta N_{\rm Total}\sim 60$) we need to set the transition scale from SRI to USR phase at $k_s\sim 10^{21}{\rm Mpc}^{-1}$, which immediate constrain the end of USR phase at $k_e\sim 10^{22}{\rm Mpc}^{-1}$. As an immediate consequence, the mass of the PBHs generated from this setup is restricted to the value $M_{\rm PBH}\propto k^{-2}_s\sim 100 {\rm gm}$. This is the new {\it no-go theorem} \cite{Choudhury:2023vuj,Choudhury:2023jlt,Choudhury:2023rks} we have proposed out of this analysis using which we arrive at the conclusion that PBH formation is strictly ruled out in single-field inflation. Now if we shift further the scale $k_s\sim 10^{6}{\rm Mpc}^{-1}$ then in that case we get $M_{\rm PBH}\sim 10^{30}{\rm gm}$, which is the solar mass. However, just by shifting the position of the transition scale, we cannot evade the {\it no-go} result as in this case, we have no successful accomplishment of inflation as we have $\Delta N_{\rm Total}\sim 20-25$.

Further, we have investigated two possibilities within the EFT framework where it is possible to evade the previously proposed {\it no-go} result for the PBH mass and can able to produce solar or sub-solar mass PBHs within the framework of EFT of single-field inflation. These possibilities are (1) Galileon inflation with previously mentioned three phases \cite{Choudhury:2023hvf,Choudhury:2023kdb,Choudhury:2023hfm,Choudhury:2023fwk,Choudhury:2024one}, (2) instead of one USR phase if we allow six USR phases \cite{Bhattacharya:2023ysp,Choudhury:2023fjs} and (3) EFT of non-singular bounce with loop effects \cite{Choudhury:2024dei} then one can evade the proposed {\it no-go theorem} on PBH mass.

\section{Discussion}

In this article, with the detailed computation of regularization, renormalization, and resummation on the one-loop corrected primordial power spectrum for scalar modes we have tried to solve the longstanding ongoing debate. To strengthen our findings we additionally come up with a strong {\it no-go theorem} on the PBH mass which helps us to arrive at the final conclusion that PBH formation in the single-field inflationary framework is ruled out. Finally, we have proposed two solutions that help us to break the mentioned {\it no-go} result which further allows us to generate solar or sub-solar mass PBHs in single-field inflation.

\vskip 0.1in
\emph{Acknowledgments:}
SC would like to thank The North American Nanohertz Observatory for Gravitational Waves (NANOGrav) collaboration and the National Academy of Sciences (NASI), Prayagraj, India, for being elected as an associate member and the member of the academy respectively. SC sincerely thanks Md. Sami and Ahaskar Karde for various useful discussions. SC acknowledge our debt to the people
belonging to various parts of the world for their generous and steady support for research in natural sciences.


\renewcommand{\leftmark}{\MakeUppercase{Bibliography}}
\phantomsection
\bibliography{Bibfile}
\bibliographystyle{utphys}


\end{document}